\documentclass[apm,reprint,amsmath,amssymb]{elsarticle}

\usepackage{graphicx}
\usepackage{bm}

\begin{document}

\title{ThFeAsN in Relation to Other Iron-Based Superconductors}

\author[um]{David J. Singh}
\ead{singhdj@missouri.edu}

\address[um]{
Department of Physics and Astronomy, University of Missouri,
Columbia, MO 65211-7010 USA}

\begin{abstract}
The electronic structure, magnetic and structural properties of the
superconductor ThFeAsN are discussed in relation to the Fe-pnictide and
Fe-chalcogenide superconductors based on results of first principles 
calculations.
The electronic structure is that of a high density of states
semimetal. It shows approximately nested hole and electron
Fermi surfaces of Fe $d$ character involving the $xz,yz$ and $xy$
orbitals. There is a strong tendency towards magnetism at the GGA
level, and this magnetism is important for describing the Fe-As
bonding, but greatly overestimates the magnetic ordering.
The lowest energy magnetic state at the GGA level is the stripe
order, and this particular order couples strongly to electrons near
$E_F$.
ThFeAsN is therefore strongly similar to the Fe-pnictide family
of superconductors, although it is a particularly anisotropic
member.
\end{abstract}

\maketitle

\section{Introduction}

Superconductivity at $T_c$=30 K was recently reported in the tetragonal
compound ThFeAsN, without doping. \cite{wang} The compound does not show
antiferromagnetic ordering, and occurs in the same crystal structure as
LaFeAsO and LaFePO. However, LaFeAsO, while tetragonal at high temperature
has a structural distortion to an orthorhombic symmetry below $\sim$160 K
and is antiferromagnetic below $\sim$145 K.
Moreover, it is not superconducting
without doping or pressure.
\cite{kamihara,cruz}
LaFePO on the other hand stays tetragonal
but only becomes superconducting at $\sim$4 K,
\cite{kamihara-p}
i.e. much lower than electron
doped LaFeAsO, which has $T_c$$\sim$26 K.
Substitution of the smaller rare earth ions, e.g. Ce, Nd, Sm and Pr for
La in electron doped LaFeAs(O,F) leads to increases in $T_c$ to over 56 K.
\cite{johnston}
Similar to LaFeAsO, the parent compounds are antiferromagnetic and
non-superconducting at ambient pressure.

From a structural point of view, ThFeAsN has a significantly shorter
$c$-axis than LaFeAsO (8.526 \AA{} vs. 8.737 \AA), but has a larger
$a$-axis (4.037 \AA{} vs. 4.030 \AA). This means that the Fe-As
planes, presumably responsible for superconductivity, have a larger in-plane
spacing, opposite to the effect of pressure on LaFeAsO and also opposite to the
effect of substituting smaller rare earth ions. The shorter $c$-axis
might suggest a more three dimensional electronic structure, but this
would depend on the details of the hopping through ThN vs. LaO.
One may also note that the behavior of the compound, in particular
superconductivity without doping, is reminiscent of some of the
Fe-chalcogenide superconductors, such as FeSe.
\cite{hsu,mizuguchi,singh-11}
The purpose of this work is to examine the electronic and related 
properties of ThFeAsN using density functional calculations.

\section{Approach}

The present calculations were done using standard density functional
calculations with the PBE generalized gradient approximation (PBE-GGA),
\cite{pbe} and the general potential linearized augmented planewave (LAPW)
method \cite{singh-book} as implemented in the WIEN2k code. \cite{wien2k}
The LAPW sphere radii were 2.4 bohr, 2.2 bohr, 2.2 bohr and 1.9 bohr for
Th, Fe, As and N, respectively. Converged basis sets consisting of
LAPW functions up to a cut-off of $R_{min}k_{max}$=8 ($R_{min}$=1.9 bohr,
the N radius) plus local orbitals were used. The results shown were
obtained in a scalar relativistic approximation, but tests were performed
including spin-orbit. We found only very small effects due to spin-orbit,
in spite of the fact that the compound contains the actinide element Th.
The reason is that the Th is very electropositive relative to N, and
as a result the Th contributions to the occupied valence states are very
small (see below).

\section{Results and Discussion}

\begin{figure*}
\includegraphics[width=\textwidth]{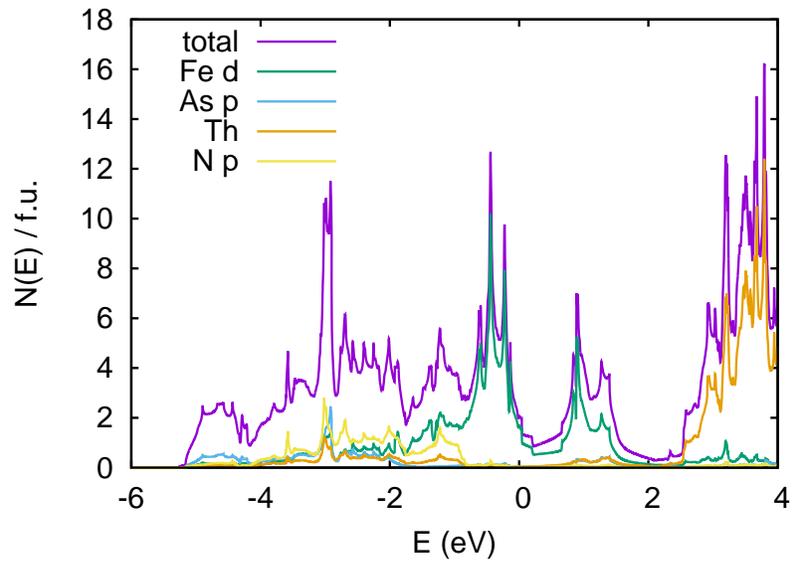}
\caption{Calculated electronic density of states of ThFeAsN and projections
on a per formula unit basis. The Fermi energy is at 0 eV.}
\label{dos}
\end{figure*}

We begin with the structure and magnetism. ThFeAsN is not reported
to show magnetic ordering. \cite{wang}
Calculations using the experimental crystal structure without magnetism
predict large forces on the As atoms. Relaxing the atomic positions
in the unit cell gives large changes in the As position.
Specifically, the As height above the Fe plane changes from
1.305 \AA{} to 1.195 \AA, leading to a decrease in Fe-As bond length of
0.058 \AA. This is far outside any reasonable error in the experimental
structure. If magnetism is included an antiferromagnetic stripe state
like that of undoped LaFeAsO is obtained. The calculated moment
in the Fe LAPW sphere is 2.01 $\mu_B$, and in this case an As position
of 1.274 \AA{} above the Fe plane is obtained, in much better accord with
experiment. This conundrum in which standard GGA calculations yield
excessive Fe-As bonding in disagreement with experimental crystal
structures without magnetism, and better structures at the expense of
greatly overestimated magnetic moments when magnetism is included is
a characteristic of the Fe-based superconductors. \cite{mazin-mag}
It has been associated with the unusual spin-fluctuations in them,
\cite{bondino}
although the subtle nature of the spin fluctuations in the Fe-based
superconductors
remains to be fully understood.
\cite{johnston,paglione,wang-mag,hosono}

We did magnetic calculations for three ordering patterns: ferromagnetic,
nearest neighbor checkerboard antiferromagnetic and stripe 
antiferromagnetic. As mentioned the lowest energy is for the stripe
pattern. With the experimental crystal structure and on a per formula
unit basis, the ferromagnetic order has an Fe moment defined by the
spin polarization in the Fe LAPW sphere of 0.36 $\mu_B$ and energy
-0.002 eV relative to the non-spin-polarized case. The nearest neighbor
checkerboard order has moment 1.81 $\mu_B$ and energy -0.059 eV, while
the stripe order has moment 2.01 $\mu_B$ and energy -0.151 eV.

The calculated energy scales for ordering and moment formation
are similar to each other. The difference in energy of the strip order
and the ferromagnetic order and that between the lowest energy magnetic
and the non-spin-polarized case are nearly the same.
Additionally,
moment size depends strongly on the particular order. 
This behavior is similar to the behavior of other Fe-based superconducting
systems in density functional calculations. It also may be taken
as indicating itinerant character of the magnetism,
\cite{singh-du,mazin-spm,kuroki}
although we note that
this characterization has been a subject of controversy in Fe-superconductors.
\cite{wang-mag,johannes}
In any case, the results show a similar behavior of ThFeAsN to the
other Fe-based superconductors.

\begin{figure*}
\includegraphics[width=\textwidth]{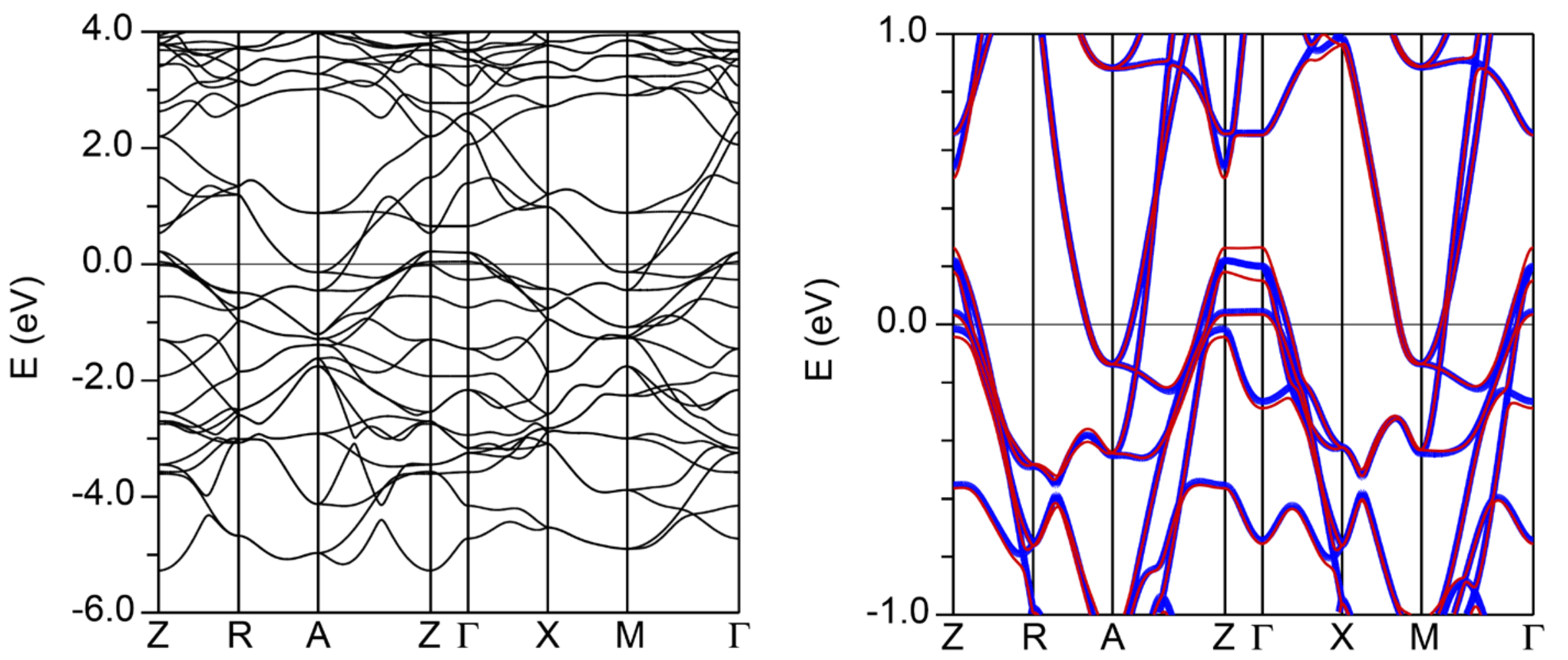}
\caption{Scalar relativistic band structure (left) and a blow up near
$E_F$ showing the scalar relativistic bands as heavy blue lines and
the bands with spin-orbit as lighter red lines (right). The
Fermi energy is at 0 eV.}
\label{bands}
\end{figure*}

We now turn to the electronic structure.
The calculated electronic density of states (DOS) is shown in Fig. \ref{dos}.
The Fermi energy, $E_F$, is located on the low energy side of a dip in
the DOS, similar to LaFeAsO and most other
Fe-based superconductors. \cite{singh-du}
This dip at an electron count of six per Fe does not correspond to
a tetrahedral crystal field (in which the 4-fold $e_g$ manifold would
be lower) but instead reflects metal-metal bonding in addition to
the interaction of Fe with the As ligands, as was discussed previously
for other Fe-based superconductors. \cite{singh-du,singh-physica}

The value of the DOS at $E_F$ is $N(E_F)$=2.02 eV$^{-1}$
per formula unit and as seen
in the projection is more than 80\% from Fe $d$ states. This places
the compound at a Stoner instability. \cite{stoner,janak}
This is the reason for the
marginal ferromagnetic instability, mentioned above.
The calculated bare specific heat coefficient from the density of states
is $\gamma_{bare}$=4.8 mJ/mol K$^2$ on a per formula unit basis.
Note, however, that in analogy
with other Fe-based superconductors there may be a substantial enhancement
due to spin-fluctuations or other correlations. 
\cite{yi2,qazilbash,tamai}

The effect of spin orbit on the density of states is small. $N(E_F)$ with
spin orbit is 2.12 eV$^{-1}$.
The Th states are unoccupied, while the As and N $p$ are below $E_F$.
The fact that the Th states are removed from the Fermi level explains the
relative insensitivity of the calculation to spin-orbit. It also means
that the Th-N layers can be regarded as (ThN)$^{+1}$ units that serve to
stabilize the structure and donate one electron to (FeAs)$^{-1}$
metallic sheets, similar to LaFeAsO.
Also because there is almost no Th character at $E_F$, the ionic
(ThN)$^{+1}$ sheets are expected to be insulating, which leads to a
rather two dimensional electronic structure for the compound, as discussed
below.

The carriers at $E_F$ are strongly coupled to the magnetism,
particularly for the stripe ordering. We obtain a strongly
reduced, $N(E_F)$=0.31 eV$^{-1}$ per formula unit for this case.
This is also a characteristic of the Fe-based superconductors.
\cite{yi,sebastian}
The nearest neighbor checkerboard
antiferromagnetic state does not gap the DOS in this way. This
order gives $N(E_F)$=4.69$^{-1}$ per formula unit, i.e. higher than
the non-magnetic state.
This shows that similar to other Fe-based superconductors, \cite{johannes}
the magnetism
involves Fe $d$ states both near and away from $E_F$ (i.e. the whole
$d$ band),
although the particular ordering
pattern realized is strongly influenced by the low energy band structure.

\begin{figure*}
\includegraphics[width=\textwidth]{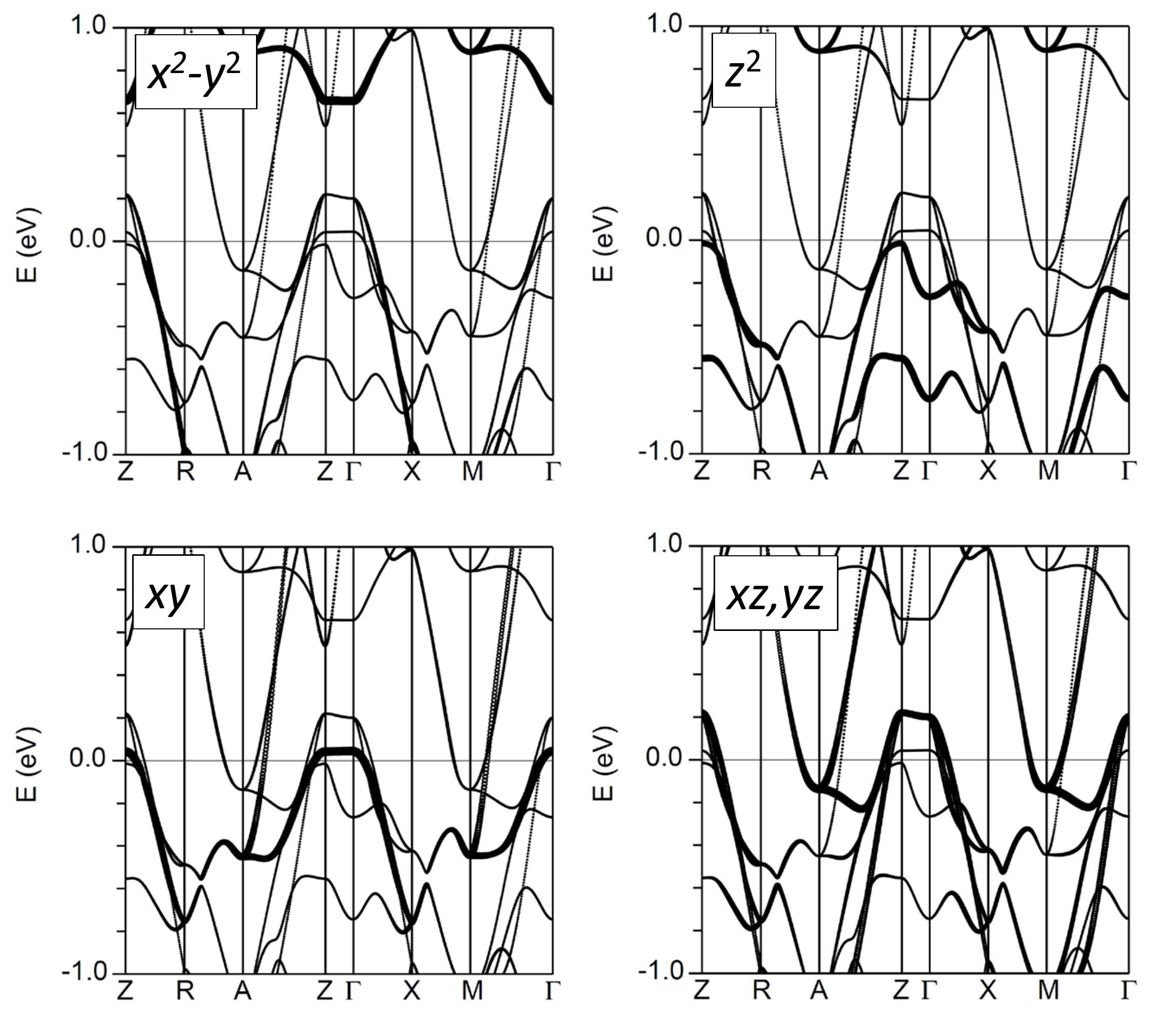}
\caption{Band structure plots emphasizing via ``fat bands" the different Fe $d$
orbital characters. The coordinates for defining
the Fe $d$ orbitals are such that $z$ is along the
$c$-axis, while the $x$ and $y$ directions are on the Fe-Fe bonds
(i.e. rotated 45$^\circ$ with respect to the two Fe atom unit cell $a$
and $b$ axes).  }
\label{band-fig}
\end{figure*}

The calculated band structure is shown in Fig. \ref{bands}, including a 
blow up of the low energy region comparing calculations with and
without spin-orbit. Plots emphasizing the different Fe $d$ orbital
characters are given in Fig. \ref{band-fig}.
The largest evident effect of spin-orbit is a splitting of the $xz,yz$ derived
bands above $E_F$
that make up the two main hole Fermi surfaces along the $\Gamma$-$Z$
direction. The effect is, however, small at $E_F$.

\begin{figure*}
\includegraphics[width=\textwidth]{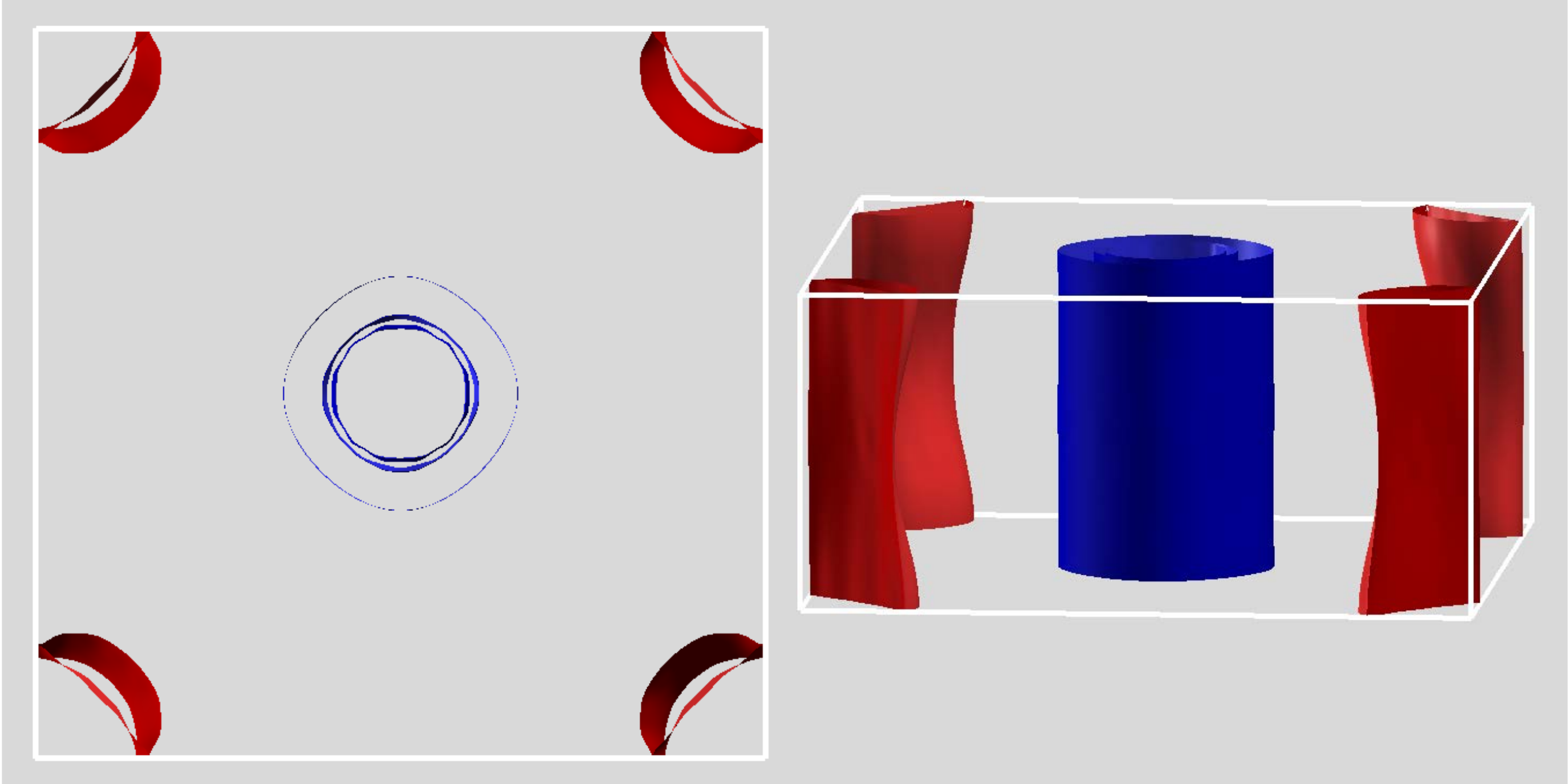}
\caption{Calculated Fermi surfaces of ThFeAsN. Hole sheets are shown
as blue and electron sheets as red.}
\label{fermi}
\end{figure*}

There are five bands crossing $E_F$: three hole bands, two of which are
of $xz,yz$ character, and one of which is of $xy$ character and
two electron bands of predominantly $xz,yz$ character hybridized with
$xy$ character. The resulting Fermi surfaces are shown in Fig. \ref{fermi}.
There are three very two dimensional and nearly circular hole cylinders
around the zone center
and two corrugated electron cylinders at the zone corner.
In addition, it is notable that there is a $z^2$ band with a maximum
just below $E_F$ at $Z$. This band differs from other bands crossing
$E_F$ in that it has considerably more dispersion in the $k_z$ direction
(see the $\Gamma$-$Z$ line in Fig. \ref{bands}).

The electron cylinders can be regarded as being
derived from the intersection of two cylinders of elliptical cross-section
with the outer lobes having more $xy$ character and the narrower (inner)
part having $xz,yz$ character.
This is similar to the other Fe-based
superconductors, including FeSe.
\cite{lebegue,singh-du,raghu,calderon,subedi,korshunov}
In particular, ThFeAsN is a high density of states, low carrier density
semimetal, with electron and hole sections separated by 
($\frac{\pi}{a},\frac{\pi}{a}$) and having similar orbital character.
This aspect of the electronic structure plays a central role in the
magnetism of Fe-based superconductors, in particular the tendency for
a stripe magnetic order that couples strongly to the carriers at $E_F$,
and the sign changing $s$-wave superconducting state that is proposed
to occur in these compounds.
\cite{mazin-spm,kuroki,christianson}

The hole surfaces contain 0.051, 0.068 and 0.153 holes per unit cell
on a both spins basis, while the electron surfaces contain 0.109
and 0.163 electrons per cell also counting both spins.
The calculated plasma frequencies are
$\Omega_{p,xx}$=$\Omega_{p,zz}$=2.31 eV and
$\Omega_{p,zz}$=0.36 eV.

In the constant scattering time approximation, the conductivity,
$\sigma\propto\Omega_p^2$. The predicted conductivity anisotropy
based on this is high, $\sigma_{xx}/\sigma_{zz}$$\sim$40.
Structural changes, e.g. by alloying, that bring the $z^2$ hole band
to $E_F$ will reduce this anisotropy.
In any case, this places ThFeAsN among the more anisotropic of the
Fe-based superconductors particularly when compared to e.g. the
ThCr$_2$Si$_2$ structure materials, such as BaFe$_2$As$_2$.
\cite{yuan,vilmercati}
This is a consequence of the fact that the (ThN)$^{+}$ sheets,
although thinner than the
(LaO)$^{+}$ sheets of LaFeAsO, are highly insulating.

\section{Summary and Conclusions}

To summarize, ThFeAsN shows several differences from the
other Fe-As superconductors, most notably the superconductivity
and lack of magnetic ordering in the absence of doping, but
nonetheless, from an electronic structure point of view it is very
similar to the other Fe-based superconductors.
The electronic structure is that of a high density of states
semimetal. It shows approximately nested hole and electron
Fermi surfaces of Fe $d$ character involving the $xz,yz$ and $xy$
orbitals. There is a strong tendency towards magnetism at the GGA
level, and this magnetism is important for describing the Fe-As
bonding, but greatly overestimates the magnetic ordering.
The lowest energy magnetic state at the GGA level is the stripe
order, and this particular order couples strongly to electrons near
$E_F$.

\section{Acknowledgements}

Support from the Department of Energy through the MAGICS center is
gratefully acknowledged.


\end{document}